\documentclass[intlimits,twoside,a4paper]{article}

\usepackage{amsmath,amssymb}
\usepackage{graphicx}
\usepackage{wrapfig}

\usepackage[T2A]{fontenc}
\usepackage[cp1251]{inputenc}
\usepackage{cmpj2}

\issue{2012}{15}{3}{33703}
\doinumber{10.5488/CMP.15.33703}

\title[Quasi-stationary states of electrons interacting with strong electromagnetic field]%
{Quasi-stationary states of electrons interacting with strong electromagnetic field in two-barrier resonance tunnel nano-structure}
\author[M.V.~Tkach, Ju.O.~Seti, O.M.~Voitsekhivska]{M.V.~Tkach\footnote{E-mail: ktf@chnu.edu.ua}\,,
Ju.O.~Seti, O.M.~Voitsekhivska}
\address{Chernivtsi National University, 2 Kotsyubinsky Str.,
58012 Chernivtsi, Ukraine}

\date{Received February 28, 2012, in final form May 5, 2012}

\authorcopyright{M.V.~Tkach, Ju.O.~Seti, O.M.~Voitsekhivska, 2012}

\begin{document}

\maketitle

\begin{abstract}
An exact solution of non-stationary Schrodinger equation is obtained for a one-dimensional movement of electrons in an electromagnetic field with arbitrary intensity and frequency. Using it, the permeability coefficient is calculated for a two-barrier resonance tunnel nano-structure placed into a high-frequency electromagnetic field. It is shown that a nano-structure contains quasi-stationary states the spectrum of which consists of the main and satellite energies. The properties of resonance and non-resonance channels of permeability are displayed.
\pacs 73.21.Fg, 73.63.Hs, 73.50.Fq, 78.67.De
\end{abstract}

\section{Introduction}

An intensive investigation of resonance tunnel structures (RTSs) is caused by their utilization in nano-devices having unique physical characteristics~\cite{Fai94,Gma01,Gio09,Lei08,Dav87,Erm00,Erm06},  and are widely used in medicine, environment monitoring and communication systems. The knowledge of the properties of RTS is urgent both from the applied and fundamental physics point of view.

The theory of ballistic and non-ballistic transport of electrons through the RTS was developed in detail mainly within the approximation of a small amplitude of high-frequency electromagnetic field~\cite{Ele03,Ele09,Pas05,Tka11,Set11,Tkc11}. Also, only linear terms over the intensity of an electric field were preserved in the Hamiltonian and wave functions. This approximation did not permit to quit the frames of the first order of the perturbation theory depending on time.

In order to clarify the effect of strong electromagnetic fields on the spectra of electrons and their tunnel through the RTS, an approximated iterating method was used for the two-level model of a nano-system~\cite{Gol99,Pas11} and a numeric method was used for a multi-level model of periodical structures~\cite{Fai97,Sac05}.

In the majority of papers, in order to use the well developed method of analytical calculations~\cite{Gol96,Ele97}, the Hamiltonian of electron-electromagnetic field interaction is written not as the one proportional to the product of electron kinetic momentum on the vector potential of the field but as a term proportional to the product of field intensity on the respective electron coordinate. The latter Hamiltonian is correct (as proven in reference~\cite{Ele08}) at the assumption that the electron interacts with the electromagnetic field inside the RTS which is not very strong. If the electromagnetic field is strong, its interaction with an electron should be taken into account in the whole space (outside the RTS too). It means that not only linear but also square terms over the kinetic momentum and vector potential must be present in a complete Hamiltonian.

In the paper, we propose an exact analytical solution of one-dimensional non-stationary Schrodinger equation obtained for the first time with the Hamiltonian of a system containing linear and square terms both over the electron kinetic momentum and vector potential of electromagnetic field. Using it, we develop a theory of transport of a mono-energetic electronic current through a two-barrier RTS placed into a high-frequency electromagnetic field with an arbitrary intensity and frequency. It is shown that the interaction between electrons and electromagnetic field is the reason why, besides the main quasi-stationary states (QSSs), there appear satellite states (QSSs) with the energies multiplied by field.

\section{Permeability coefficient for a two-barrier RTS placed into a  high-fre\-quen\-cy electromagnetic field}

\begin{wrapfigure}{o}{0.4\textwidth}
\centerline{
\includegraphics[width=0.3\textwidth]{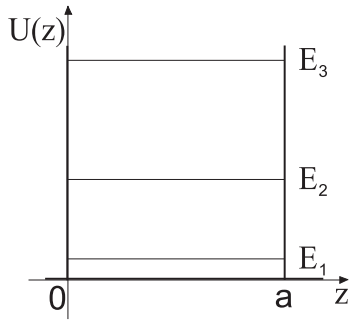}
}
\caption{Energy scheme for an electron in two-barrier RTS with $\delta$-like potential barriers.} \label{fig1}
\end{wrapfigure}

We study the symmetric two-barrier RTS (figure~\ref{fig1}) in a homogeneous high-frequency electromagnetic field: $\mathcal{E}(t)=2 \mathcal{E} \cos(\omega t)$ with an arbitrary electric field intensity $\mathcal{E}$ and frequency $\omega$. The mono-energetic electronic current with the energy $E$ moving perpendicularly to the planes of RTS gets in it from the left hand side.

The electron wave function has to satisfy a complete one-dimensional Schrodinger equation
\begin{equation}
\label{eq1}
\ri \hbar \frac{\partial \Psi(z, t)}{\partial t}=\left[\frac{1}{2 m} \left(\hat{p}_{z}-\frac{e}{c} A_{z}\right)^{2}+U(z)\right]\Psi(z, t).
\end{equation}
Using the known expressions for a kinetic momentum operator $\hat{p}_{z}$ and vector potential $A_{z}$ written in Coulomb calibration, equation (\ref{eq1}) takes the form
\begin{equation}
\label{eq2}
\ri \hbar \frac{\partial \Psi(z, t)}{\partial t}=\left[H_\mathrm{e} + U(z) + H_\mathrm{int}\right] \Psi(z,t).
\end{equation}

The complete Hamiltonian contains an electron kinetic energy operator
\begin{equation}
\label{eq3}
H_\mathrm{e} = -\frac{\hbar^{2}}{2 m}\frac{\partial^{2}}{\partial z^{2}}\,,
\end{equation}
the potential energy of electron in two-barrier RTS written within the typical $\delta$-barrier approximation, references~\cite{Ele03,Ele09,Pas05}
\begin{equation}
\label{eq4}
U(z)=U \Delta \left[\delta(z)+\delta(z-a)\right],
\end{equation}
and the potential energy of interaction between the electron and electromagnetic field
\begin{equation}
\label{eq5}
H_\mathrm{int}=-\frac{2 \ri e \mathcal{E} \hbar}{m \omega}\sin(\omega t)\frac{\partial}{\partial z} + \frac{2 (e \mathcal{E})^{2}}{m \omega^{2}} \sin^{2}(\omega t),
\end{equation}
where $e$, $m$ are the electron charge and mass; $U$, $\Delta$ are the height and width of potential barriers; $a$ is the width of potential well.

The equation (\ref{eq2}) has two exact linearly independent solutions for all parts of a nano-structure
\begin{equation}
\label{eq6}
\psi^{\pm}(E,\omega, z, t)=\exp \left\{\pm \ri k_{0}\left[z+\frac{2 e\mathcal{E}}{m \omega^{2}} \cos(\omega t)\right]-\frac{\ri}{\hbar} \left[E+\frac{(\re \mathcal{E})^{2}}{m \omega^{2}} \left(1-\frac{\sin(2 \omega t)}{2 \omega t}\right)\right]t \right\},
\end{equation}
describing the incident and the reflected waves with quasi-momentum $k_{0} = \hbar^{-1} \sqrt{2 m E}$.

A complete wave function, being a linear combination of both solutions (\ref{eq6}), taking into account the expansion of all periodical functions into Fourier range and considering the super positions with all electromagnetic field harmonics, is written as follows:
\begin{equation}
\label{eq7}
\Psi(E, \omega, z, t)=
\sum\limits_{p = -\infty}^{+\infty}\left[ \psi^{+}(E + p \Omega, \omega, z, t)+ \psi^{-}(E + p \Omega, \omega, z, t)\right],
\end{equation}
where
%% \label{eq8}
\begin{eqnarray}
\psi^{\pm}(E + p \Omega, \omega, z, t)&=&\re^{\pm \ri k_{p} z-\frac{\ri}{\hbar}(E + p \Omega)t} \sum \limits_{n_1 = -\infty}^{+\infty} f_{n_{1}}(\pm k_{p},\omega, t) \re^{- \ri n_{1} \omega t}
\nonumber\\
&&{}\times \left\{B^{\pm}_{0, p} \theta(-z) + B^{\pm}_{1, p}[\theta(z) - \theta(z-a)] + B^{\pm}_{2, p} \theta(z-a)\right\},
\label{multi-line}
\end{eqnarray}
\begin{equation}
\label{eq9}
f_{n_{1}}(\pm k_{p},\omega, t)=\frac{1}{2 \pi} \int \limits_{-\pi}^{\pi} \exp \left\{\ri n_{1} \xi \pm 4\ri \alpha \beta a k_{p} \cos \xi - 2 \ri \alpha \beta^{2} \xi \left[1-\frac{\sin(2 \xi)}{2 \xi}\right] \right\}\rd \xi.
\end{equation}
Here, convenient denotations are
\begin{equation}
\label{eq10}
\Omega=\hbar \omega;\qquad \alpha=\frac{\hbar^{2} k_{a}^{2}}{2m \Omega}; \qquad \beta=\frac{U_{a}}{\Omega}; \qquad U_{a}=e \mathcal{E} a; \qquad k_{p}=\hbar^{-1} \sqrt{2m(E+p\Omega)}; \qquad k_{a}=a^{-1}
\end{equation}
with the evident physical sense: $\alpha$ is the kinetic energy of an electron with quasi-momentum $k_a$, $\beta$ is the potential energy of an electron interacting with the electromagnetic field written in the units of electromagnetic field energy $\Omega$.

All unknown coefficients: $B^{\pm}_{(0,1,2),p}$ are obtained from the conditions of a wave function and its density of current continuity at RTS interfaces at any moment of time $t$:
%% \label{eq11}
\begin{alignat}{4}
&\Psi(E, \omega,+\eta, t)=\Psi(E, \omega,-\eta, t)  \qquad  (\eta \rightarrow 0)\, ;  \nonumber \\[2ex]
& \frac{\partial}{\partial z}\Psi(E, \omega, z, t)\bigg|_{z=+\eta} - \frac{\partial}{\partial z}\Psi(E, \omega, z, t)\bigg|_{z=-\eta} = \frac{U \Delta k_{a}^{2}}{\alpha \Omega} \Psi(E, \omega, 0, t)\, ;  \nonumber \\[2ex]
&\Psi(E, \omega,a +\eta, t)=\Psi(E, \omega, a -\eta, t)\, ;  \nonumber \\[2ex]
& \frac{\partial}{\partial z}\Psi(E, \omega, z, t)\bigg|_{z=a+\eta} - \frac{\partial}{\partial z}\Psi(E, \omega, z, t)\bigg|_{z=a-\eta} =  \frac{U \Delta k_{a}^{2}}{\alpha \Omega} \Psi(E, \omega, a, t) . \label{multi-line_2}
\end{alignat}
Also, $B^{+}_{0,p\neq 0}=B^{-}_{2,p}=0$ because the mono-energetic electronic current gets in RTS only along the main channel ($p=0$) and there are no incident currents along the other channels ($p\neq 0$).

The system of equations (11) contains an infinite number of equations due to an infinite number of harmonics. Performing the calculations, it can be confined by the sufficient arbitrarily big but finite number of positive $N^+$ and negative $N^-$ harmonics which are limited by the number of open channels fixed by an obvious condition: $N^{-} < [E/\Omega]$. Thus, in the finite system of $4 (N^{-}+N^{+}+1)$ equations respectively the same number of coefficients is obtained.

The law of conservation of a complete density of current through all the open channels should be fulfilled
\begin{equation}
\label{eq12}
J^{+}(E, \omega, z=0) = \sum\limits_{p = -N^{-}}^{N^{+}}\left[J^{-}(E+p\Omega, \omega, z=0)+J^{+}(E+p\Omega, \omega, z=a)\right].
\end{equation}
Thus, calculating the densities of forward $J^+$ and backward $J^-$ electronic currents getting in RTS ($z=0$) and coming out of  RTS ($z=a$), according to reference~\cite{Lan81}, the permeability coefficient is expressed through partial terms
\begin{equation}
\label{eq13}
D(E, \omega) = \sum\limits_{p = -N^{-}}^{N^{+}}J^{+}(E+p\Omega, \omega, a)\left[J^{+}(E, \omega, 0)\right]^{-1}=\sum\limits_{p = -N^{-}}^{N^{+}}D_{p}(E+p\Omega, \omega).
\end{equation}

Using it, one can obtain the resonance energies and widths of the main and arbitrary number of satellite QSSs for the electrons interacting with a high-frequency electromagnetic field. The developed theory proves that this interaction causes renormalization of ``pure'' electron QSSs and, besides, the appearance of satellite QSSs corresponding to all possible electromagnetic field harmonics. Consequently, the respective maxima of permeability coefficient can be observed when a satellite QSS of a certain electron state should resonate with the main or satellite state of the other QSS, producing complex QSSs.

\section{Properties of quasi-stationary spectrum of electron-electro\-mag\-ne\-tic field system in a two-barrier RTS}

It is well known, references~\cite{Pas11, Tka09}, that when there is no electromagnetic field, a quasi-stationary electron spectrum is characterized  by resonance energies $E_n$, with the magnitudes determined by the maxima of permeability coefficient $D$ in energy scale. Similarly, a quasi-stationary spectrum of electrons interacting with electromagnetic field is defined by the positions of all maxima of the permeability coefficient. In this spectrum one can see the satellite QSSs with the energies: $E_{n(p)} = E_{n} + p \Omega$, arising near each of the QSSs with ``pure'' resonance energies $E_n$.

The calculations of permeability coefficient and the energies of QSSs were performed for In$_{0.52}$Al$_{0.48}$As/ In$_{0.53}$Ga$_{0.47}$As two-barrier RTS, often studied experimentally~\cite{Fai94,Gma01,Gio09}. The physical parameters are: $m = 0.043~m_e$ ($m_e$ is the pure electron mass), $U=516$~meV and geometrical ones: $a = 24$~nm, $\Delta = 12$~nm.

\begin{figure}[!t]
\centerline{
\includegraphics[width=0.65\textwidth]{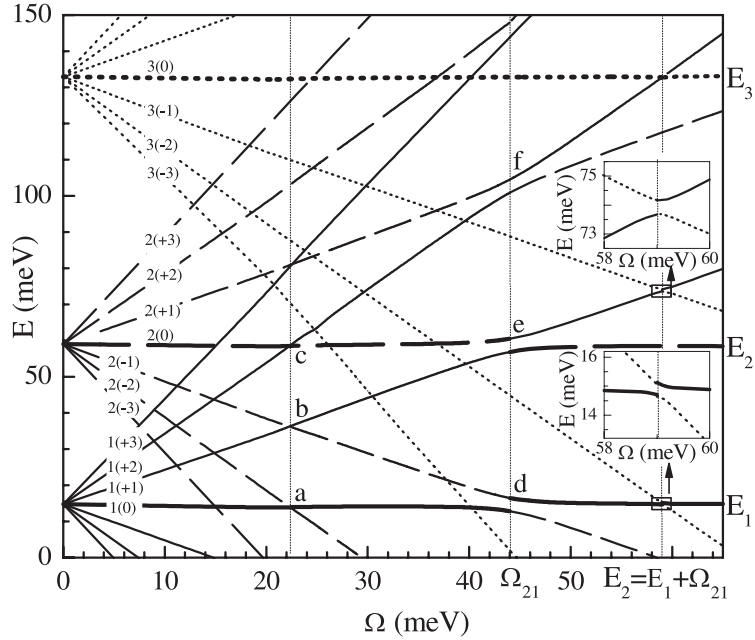}
}
\caption{The energies of main and satellite QSSs as functions of electromagnetic field energy~$\Omega$ at $U_{a} = 10$~meV in two-barrier RTS with $a = 24$~nm, $\Delta = 12$~nm.} \label{fig2}
\end{figure}

In figure~\ref{fig2}, the results of numeric calculations of the energies of the main and satellite QSSs are shown as functions of the electromagnetic field energy $\Omega$ at a fixed shift energy ($U_{a} = 10$~meV) arising due to the electric intensity $\mathcal{E}$. From the figure it is clear that at certain field energies, the energies of main and satellite QSSs coincide in pairs, which means that these states degenerate (points a, b, c). Anti-crossings are observed for the other field energies (d, e, f).

One can see that the first series of anti-crossings is located in the vicinity of the field energy: $\Omega = \Omega_{21} = E_{2}-E_{1}$ corresponding to the difference between the resonance energies of the second and the first QSS. The second series of anti-crossings is located in the vicinity of the field energy: $\Omega = E_{2} = E_{1}+\Omega_{21}$. It is clear that the series of anti-crossings are located in the vicinity of the field energies corresponding to the energies of all possible harmonics and their combinations (super positions) with the energies of ``pure'' QSSs.

For the field energies where there are still no anti-crossings, the  energies of QSSs can be consistently characterized by two indexes: $E_{n(p)}$, where $p=0$ is the main state and $p = \pm 1, \pm 2, \ldots$ is the number of positive ($+$) or negative ($-$) satellite states for $n$-th QSS. The same indexes are correct for the points of degeneration (a, b, c at figure~\ref{fig2}). However, a more detailed indexation should be introduced in the vicinity of field energies where anti-crossings occur.  The indexes should show how  the energy of the state $n(p)$ transforms into the energy of the state $n'(p')$ for each of the both complex states producing an anti-crossing at an increase of the field energy. So, the energies of the first anti-crossing pair (d at figure~\ref{fig2}) are written as: $E_{1(0); 2(-1)}$ (lower one) and $E_{2(-1); 1(0)}$ (upper one). Here, the first pair of numbers means from what state this complex state started, and the second pair means into what state it transferred at an increase of the field energy. The proposed indexes consistently characterize all the double complex states. Also, it is evident that in the vicinity of all series of anti-crossings the rule $n+p = n'+p'$ is fulfilled. It means that the sum of indexes for each complex QSS in an anti-crossing pair is the same. The sizes of all anti-crossings in certain series (the first at $\Omega = \Omega_{21}$, the second at $\Omega = E_{2} = E_{1}+\Omega_{21}$) and the distances between its neighbouring anti-crossings are the same too. The distance between the neighbouring anti-crossings of the first series is, naturally, smaller than that of the second one. The sizes of  anti-crossings of the first series are much bigger than those of the second series.

\begin{figure}[!b]
\centerline{
\includegraphics[width=0.9\textwidth]{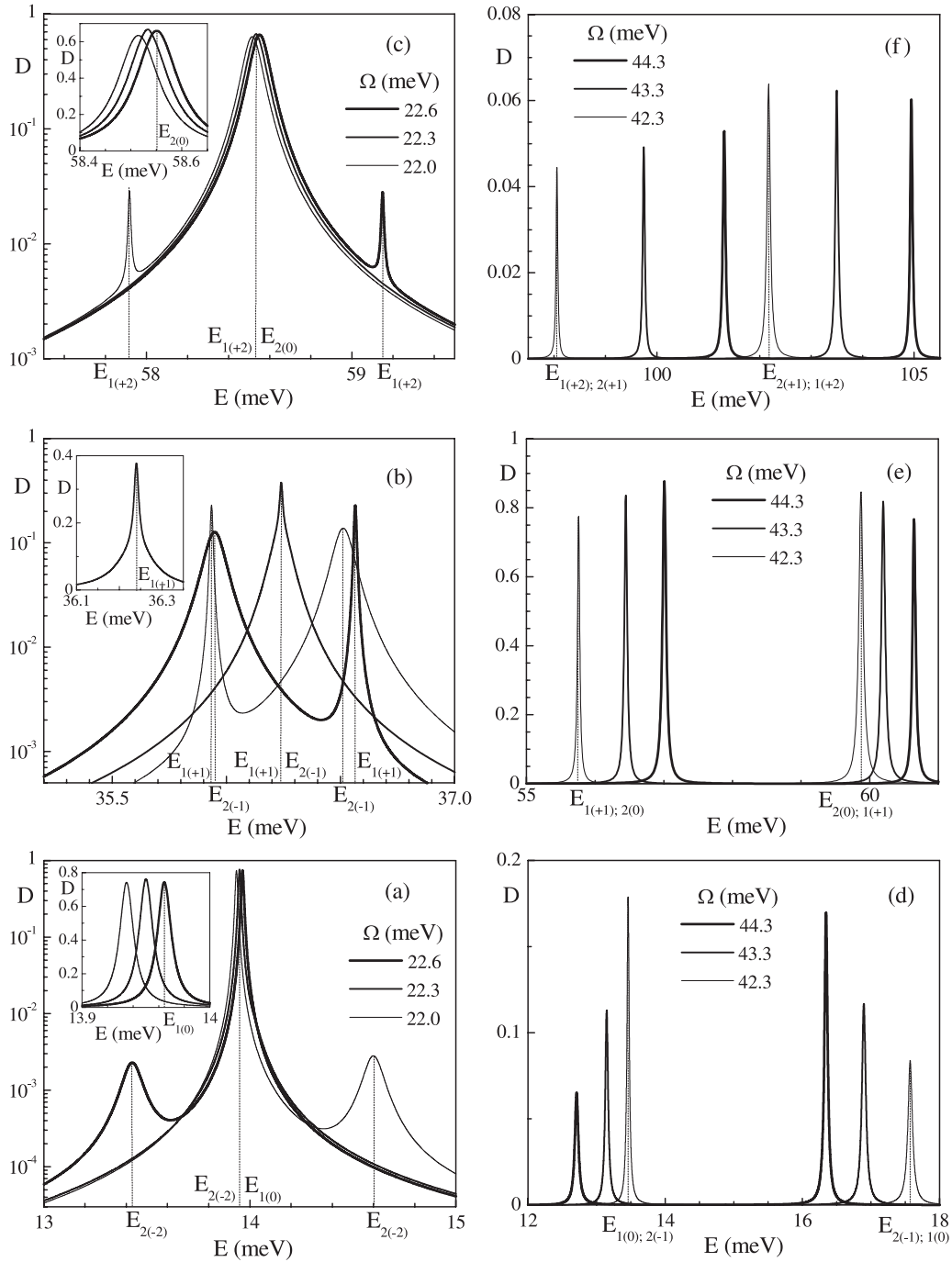}
}
\caption{Dependence of permeability coefficient $D$ on electron energy $E$ at a fixed electromagnetic field energy $\Omega$ in the vicinity of degeneration points [(a), (b), (c)] and in the vicinity of anti-crossings [(d), (e), (f)] in two-barrier RTS with $a = 24$~nm, $\Delta = 12$~nm.} \label{fig3}
\end{figure}

In figure~\ref{fig3}, the dependencies of permeability coefficient $D$ on the electron energy $E$ are presented in the vicinity of degeneration (points a, b, c) and in the vicinity of anti-crossings (d, e, f) for three different magnitudes of the field energy written in the figure. The latter are chosen in such a way that the points of degeneration or minimal magnitudes of anti-crossings are enveloped from both sides. We should also note that the characters near the points of degeneration (a, b, c) and near the anti-crossings (d, e, f) in figure~\ref{fig2} correspond to those in figure~\ref{fig3}.

In figure~\ref{fig3}~(a),~(c), the permeability coefficient is presented as a function of field energy in the vicinity of points of degeneration at the cross of the energy of the first main state $E_{1(0)}$ with the energy of the second negative satellite $E_{2(-2)}$ of the second main state $E_{2(0)}$, figure~\ref{fig3}~(a) and at the cross of the second main state energy $E_{2(0)}$ with the energy of the second positive satellite $E_{1(+2)}$ of the first main state $E_{1(0)}$, figure~\ref{fig3}~(c). The both figures~\ref{fig3}~(a), (c) prove that in accordance with the figure~\ref{fig2}, an increase of field energy $\Omega$ weakly shifts the peaks of the main states into the region of smaller or bigger energies and almost does not change the magnitudes of their maxima ($D_{1(0)} \approx 0.76$, $D_{2(0)} \approx 0.67$). Herein, the satellite peak with a small permeability ($D_{2(-2)}\approx 0.002$) near the first main state essentially shifts into the region of smaller energies [figure~\ref{fig3}~(a)] while the satellite peak with the small permeability ($D_{1(+2)} \approx 0.029$) near the second main state essentially shifts into the region of bigger energies [figure~\ref{fig3}~(c)]. Consequently, the permeability of the main channels  prevails while the permeability of satellite channels is negligibly small in the vicinity of degeneration points of the main QSSs with satellite channels.

The permeability of satellite channels decreases at an increasing number of the respective harmonics $|p|$ and remains negligibly small till the pair of satellite QSSs approaches rather close to the point of degeneration (b at figure~\ref{fig2}) at the increasing field energy. An example of the behaviour of a pair of  satellite QSSs as a function of field energy is shown in figure~\ref{fig3}~(b). It proves that at bigger $\Omega$, the first positive satellite $E_{1(+1)}$ of the first main state that shifts into the region of bigger energies, passes the point of degeneration almost without changing its small width. The first negative satellite $E_{2(-1)}$ of the second main state moves in the opposite direction. It also  does not almost change its width which is bigger than the width of the first peak. It is obvious that the permeability of both channels (till the moment of degeneration) is not essential ($D_{1(+1)} \approx 0.23$, $D_{2(-1)} \approx 0.14$) but  already not negligibly small. The permeability of a common channel in the point of degeneration almost coincides with the sum of permeabilities of the both composing channels ($D_{1(+1)}+D_{2(-1)} \approx 0.38$).

We have already noted that in the vicinity of the field energies $\Omega_{21}$ close to the difference of resonance energies, a series of anti-crossings is observed. They are produced by double complexes of QSSs which are the super-positions of a pair of main states with the satellites of the other states [figure~\ref{fig3}~(d),~(e)] or a pair of satellites of different main QSSs [figure~\ref{fig3}~(f)].

Figures~\ref{fig2} and \ref{fig3} prove that the distances between the energies of both states in each anti-crossing decrease at first, approaching their minima and, then, increase at $\Omega$ increasing. Herein, an essential permeability of the main channels produced by double complexes of the main and satellite QSSs [figure~\ref{fig3}~(d), (e)] decreases at first and increases for the satellite channels. At $\Omega \simeq \Omega_{21}$, the permeabilities of the both channels become the same. Then, the permeability of satellite channels decreases and that of the main channels increases. The permeabilities of double satellite channels [figure~\ref{fig3}~(f)] are small and nearly the same in the vicinity of an anti-crossing.

\newpage
\section{Conclusions}

\begin{enumerate}
\item
Using an exact solution of non-stationary one-dimensional Schrodinger equation for electrons interacting with a high-frequency electromagnetic field and passing through the two-barrier RTS we established the existence of resonance and non-resonance channels of permeability. It is proven that there are observed the main satellite and double complexes of QSSs producing different channels of two-barrier RTS permeability.

\item In the vicinity of  electromagnetic field energies that resonate with the difference of the energies of the main QSSs, a series of anti-crossings arise both between the main states with the field satellites of the other main states and between the satellites of different states with each other.

\item As far as the permeability of the both channels in a pair complex, producing an anti-crossing of the main and satellite states, is rather big, it can essentially effect the operation of nano-lasers and nano-detectors the basic element of which is a nano-RTS.
\end{enumerate}

%% Type in your references using {thebibliography} environment
%% or create them from your bibtex database using cmp.bst style.

%\bibliographystyle{cmp}
%\bibliography{mybibdb}

\newpage
\ukrainianpart

\title{Квазістаціонарні стани електронів, що взаємодіють з сильним електромагнітним полем у двобар'єрній наносистемі}
\author{М.В.Ткач, Ю.О.Сеті, О.М.Войцехівська}
\address{Чернівецький національний університет ім.Ю.~Федьковича, \\Україна, 58012 Чернівці, вул. Коцюбинського, 2}

\makeukrtitle

\begin{abstract}
\tolerance=3000%
Знайдено точний розв'язок нестаціонарного рівняння Шредінґера для одновимірного руху електронів у електромагнітному полі довільних величин напруженості та частоти. На цій основі виконано розрахунок коефіцієнта електронної прозорості двобар'єрної резонансно-тунельної наноструктури у високочастотному електромагнітному полі. Показано, що досліджувана система має квазістаціонарні стани, спектр яких містить основні та сателітні енергії. Виявлено властивості резонансних і нерезонансних каналів прозорості двобар'єрної наносистеми.
\keywords резонансно-тунельна наноструктура, коефіцієнт прозорості, електромагнітне поле

\end{abstract}

\end{document}